\documentstyle[prl,aps,eqsecnum,floats,epsf,tighten]{revtex}

\def\met{\mbox{${\hbox{$E$\kern-0.6em\lower-.1ex\hbox{/}}}_T$}}
\def\kt{{\it k}$_{\perp}\:$}

\def\2{{\it k}$_{\perp}$}
\def\kti{{\it k}$_{\perp}$}
\def\pt{{\it p}$_{T}\:$}
\def\et{{\it E}$_{T}\:$}
\newcommand{\Rsep}{\mbox{${\mathcal{R}}_{\rm sep}$}}

\begin{document}

\draft

\title{The Inclusive Jet Cross Section in 
$p\overline{p}$ Collisions at $\sqrt{s}=1.8$~TeV 
using the \kt Algorithm}

%
\author{                                                                      
V.M.~Abazov,$^{23}$                                                           
B.~Abbott,$^{57}$                                                             
A.~Abdesselam,$^{11}$                                                         
M.~Abolins,$^{50}$                                                            
V.~Abramov,$^{26}$                                                            
B.S.~Acharya,$^{17}$                                                          
D.L.~Adams,$^{59}$                                                            
M.~Adams,$^{37}$                                                              
S.N.~Ahmed,$^{21}$                                                            
G.D.~Alexeev,$^{23}$
% ...
\\                                                                            
\vskip 0.30cm                                                                 
\centerline{(D\O\ Collaboration)}                                             
\vskip 0.30cm                                                                 
}                                                                             
\address{                                                                     
\centerline{$^{1}$Universidad de Buenos Aires, Buenos Aires, Argentina}       
\centerline{$^{2}$LAFEX, Centro Brasileiro de Pesquisas F{\'\i}sicas,         
                  Rio de Janeiro, Brazil}                                     
\centerline{$^{3}$Universidade do Estado do Rio de Janeiro,                   
                  Rio de Janeiro, Brazil}                                     
\centerline{$^{4}$Institute of High Energy Physics, Beijing,                  
                  People's Republic of China}                                 
\centerline{$^{5}$Universidad de los Andes, Bogot\'{a}, Colombia}             
\centerline{$^{6}$Charles University, Center for Particle Physics,            
                  Prague, Czech Republic}                                     
\centerline{$^{7}$Institute of Physics, Academy of Sciences, Center           
                  for Particle Physics, Prague, Czech Republic}               
\centerline{$^{8}$Universidad San Francisco de Quito, Quito, Ecuador}         
\centerline{$^{9}$Institut des Sciences Nucl\'eaires, IN2P3-CNRS,             
                  Universite de Grenoble 1, Grenoble, France}                 
\centerline{$^{10}$CPPM, IN2P3-CNRS, Universit\'e de la M\'editerran\'ee,     
                  Marseille, France}                                          
\centerline{$^{11}$Laboratoire de l'Acc\'el\'erateur Lin\'eaire,              
                  IN2P3-CNRS, Orsay, France}                                  
\centerline{$^{12}$LPNHE, Universit\'es Paris VI and VII, IN2P3-CNRS,         
                  Paris, France}                                              
\centerline{$^{13}$DAPNIA/Service de Physique des Particules, CEA, Saclay,    
                  France}                                                     
\centerline{$^{14}$Universit{\"a}t Mainz, Institut f{\"u}r Physik,            
                  Mainz, Germany}                                             
\centerline{$^{15}$Panjab University, Chandigarh, India}                      
\centerline{$^{16}$Delhi University, Delhi, India}                            
\centerline{$^{17}$Tata Institute of Fundamental Research, Mumbai, India}     
\centerline{$^{18}$Seoul National University, Seoul, Korea}                   
\centerline{$^{19}$CINVESTAV, Mexico City, Mexico}                            
\centerline{$^{20}$FOM-Institute NIKHEF and University of                     
                  Amsterdam/NIKHEF, Amsterdam, The Netherlands}               
\centerline{$^{21}$University of Nijmegen/NIKHEF, Nijmegen, The               
                  Netherlands}                                                
\centerline{$^{22}$Institute of Nuclear Physics, Krak\'ow, Poland}            
\centerline{$^{23}$Joint Institute for Nuclear Research, Dubna, Russia}       
\centerline{$^{24}$Institute for Theoretical and Experimental Physics,        
                   Moscow, Russia}                                            
\centerline{$^{25}$Moscow State University, Moscow, Russia}                   
\centerline{$^{26}$Institute for High Energy Physics, Protvino, Russia}       
\centerline{$^{27}$Lancaster University, Lancaster, United Kingdom}           
\centerline{$^{28}$Imperial College, London, United Kingdom}                  
\centerline{$^{29}$University of Arizona, Tucson, Arizona 85721}              
\centerline{$^{30}$Lawrence Berkeley National Laboratory and University of    
                  California, Berkeley, California 94720}                     
\centerline{$^{31}$University of California, Davis, California 95616}         
\centerline{$^{32}$California State University, Fresno, California 93740}     
\centerline{$^{33}$University of California, Irvine, California 92697}        
\centerline{$^{34}$University of California, Riverside, California 92521}     
\centerline{$^{35}$Florida State University, Tallahassee, Florida 32306}      
\centerline{$^{36}$Fermi National Accelerator Laboratory, Batavia,            
                   Illinois 60510}                                            
\centerline{$^{37}$University of Illinois at Chicago, Chicago,                
                   Illinois 60607}                                            
\centerline{$^{38}$Northern Illinois University, DeKalb, Illinois 60115}      
\centerline{$^{39}$Northwestern University, Evanston, Illinois 60208}         
\centerline{$^{40}$Indiana University, Bloomington, Indiana 47405}            
\centerline{$^{41}$University of Notre Dame, Notre Dame, Indiana 46556}       
\centerline{$^{42}$Iowa State University, Ames, Iowa 50011}                   
\centerline{$^{43}$University of Kansas, Lawrence, Kansas 66045}              
\centerline{$^{44}$Kansas State University, Manhattan, Kansas 66506}          
\centerline{$^{45}$Louisiana Tech University, Ruston, Louisiana 71272}        
\centerline{$^{46}$University of Maryland, College Park, Maryland 20742}      
\centerline{$^{47}$Boston University, Boston, Massachusetts 02215}            
\centerline{$^{48}$Northeastern University, Boston, Massachusetts 02115}      
\centerline{$^{49}$University of Michigan, Ann Arbor, Michigan 48109}         
\centerline{$^{50}$Michigan State University, East Lansing, Michigan 48824}   
\centerline{$^{51}$University of Nebraska, Lincoln, Nebraska 68588}           
\centerline{$^{52}$Columbia University, New York, New York 10027}             
\centerline{$^{53}$University of Rochester, Rochester, New York 14627}        
\centerline{$^{54}$State University of New York, Stony Brook,                 
                   New York 11794}                                            
\centerline{$^{55}$Brookhaven National Laboratory, Upton, New York 11973}     
\centerline{$^{56}$Langston University, Langston, Oklahoma 73050}             
\centerline{$^{57}$University of Oklahoma, Norman, Oklahoma 73019}            
\centerline{$^{58}$Brown University, Providence, Rhode Island 02912}          
\centerline{$^{59}$University of Texas, Arlington, Texas 76019}               
\centerline{$^{60}$Texas A\&M University, College Station, Texas 77843}       
\centerline{$^{61}$Rice University, Houston, Texas 77005}                     
\centerline{$^{62}$University of Virginia, Charlottesville, Virginia 22901}   
\centerline{$^{63}$University of Washington, Seattle, Washington 98195}       
}                                                                             

\maketitle
\vskip 0.5cm

\begin{abstract}
\noindent

The central inclusive jet cross section has been measured using a
successive-combination algorithm for reconstruction of jets. The
measurement uses $87.3$~pb$^{-1}$ of data collected with the D\O\
detector at the Fermilab Tevatron $p\overline{p}$ Collider during
1994--1995. The cross section, reported as a function of transverse
momentum (\pt $>60$ GeV) in the central region of pseudorapidity
($|\eta|<0.5$), exhibits reasonable agreement with next-to-leading
order QCD predictions, except at low $p_T$ where the agreement is
marginal.

\end{abstract}

\twocolumn
\narrowtext

Jet production in hadronic collisions is understood within the
framework of Quantum Chromodynamics (QCD) as a hard scattering of
constituent partons (quarks and gluons), that, having undergone the
interaction, manifest themselves as showers of collimated particles
called jets. Jet finding algorithms associate clusters of these
particles into jets so that the kinematic properties of the
hard-scattered partons can be inferred and thereby compared to
predictions from perturbative QCD (pQCD).

Historically, only cone algorithms have been used to reconstruct jets
at hadron colliders~\cite{snowmass}.  Although well-suited to the
understanding of the experimental systematics present in the complex
environment of hadron colliders, the cone algorithms used in previous
measurements by the Fermilab Tevatron
experiments~\cite{d0_jets_prd,cdf_jets_prd} present several
difficulties: an arbitrary procedure must be implemented to split and
merge overlapping calorimeter cones, an ad-hoc parameter,
\Rsep~\cite{rsep}, is required to accommodate the differences between
jet definitions at the parton and detector levels, and improved
theoretical predictions calculated at the
next-to-next-to-leading-order (NNLO) in pQCD are not infrared safe,
but exhibit sensitivity to soft radiation~\cite{soft_rad}.

A second class of jet algorithms, which does not suffer from these
shortcomings, has been developed by several
groups~\cite{catani93,catani92,ellis93}.  These recombination
algorithms successively merge pairs of nearby objects (partons,
particles, or calorimeter towers) in order of increasing relative
transverse momentum.  A single parameter, $D$, which approximately
characterizes the size of the resulting jets, determines when this
merging stops.  No splitting or merging is involved because each
object is uniquely assigned to a jet.  There is no need to introduce
any ad-hoc parameters, because the same algorithm is applied at the
theoretical and experimental level.  Furthermore, by design,
clustering algorithms are infrared and collinear safe to all orders of
calculation.  In this Letter, we present the first measurement of the
inclusive jet cross section using the \kt
algorithm~\cite{catani93,catani92,ellis93,rob_prd} to reconstruct jets
at the $\sqrt{s}=1.8$ TeV Tevatron proton-antiproton collider.

The differential jet cross section was measured in bins of \pt\ and
pseudorapidity, $\eta \equiv -{\rm ln}[{\rm tan}(\theta/2)]$, where
$\theta$ is the polar angle relative to the $z$ axis pointing in the
proton beam direction.  The \kt algorithm implemented at D\O
~\cite{rob_prd} is based on the clustering algorithm suggested in
Ref.~\cite{ellis93}.  The algorithm starts with a list of objects.
For each object with transverse momentum $p_{T,i}$, we define
$d_{ii}=p_{T,i}^2$, and for each pair of objects,
$d_{ij}=\text{min}(p_{T,i}^2,p_{T,j}^2)\,(\Delta R_{i,j})^2/D^2$,
where $D$ is the free parameter of the algorithm and $(\Delta
R_{i,j})^2=(\Delta\phi_{ij})^2+(\Delta\eta_{ij})^2$ is the square of
their angular separation.  If the minimum of all $d_{ii}$ and $d_{ij}$
is a $d_{ij}$, then the objects $i$ and $j$ are combined, becoming the
merged four-vector $(E_i+E_j,\,\vec{p}_i+\vec{p}_j)$.  If the minimum
is a $d_{ii}$, the object $i$ is defined as a jet and removed from
subsequent iterations.  This procedure is repeated until all objects
are combined into jets.  Thus \kt\ jets do not have to include all
objects in a cone of radius $D$, and they may include objects outside
of this cone.

The primary tool for jet detection at D\O\ is the liquid-argon/uranium
calorimeter~\cite{d0calor}, which has nearly full solid-angle coverage
for $|\eta|<4.1$.  The first stage (hardware) trigger selected
inelastic collisions as defined by signal coincidence in the
hodoscopes located near the beam axis on both sides of the interaction
region. The next stage required energy deposition in any
$\Delta\eta\times\Delta\phi=0.8\times 1.6$ region of the calorimeter
corresponding to a transverse energy ({\it E}$_{T}$) above a preset
threshold. Selected events were digitized and sent to an array of
processors. At this stage jet candidates were reconstructed with a
cone algorithm (with radius $R\equiv
[(\Delta\eta)^2+(\Delta\phi)^2]^{\frac{1}{2}}=0.7$), and the event was
recorded if any jet \et exceeded a specified threshold. Jet \et
thresholds of $30$, $50$, $85$, and $115$ GeV accumulated integrated
luminosities of $0.34$, $4.46$, $51.5$, and $87.3$ pb$^{-1}$,
respectively~\cite{thesis}.

Jets were reconstructed offline using the \kt algorithm, with
$D=1.0$. This value of $D$ was chosen because, at
next-to-leading-order (NLO), it produces a theoretical cross section
that is essentially identical to the cone prediction for
$R=0.7$~\cite{ellis93}, which D\O\ used in its previous publications
on jet production~\cite{d0_jets_prd}.  The vertices of the events were
reconstructed using the central tracking system~\cite{d0calor}. A
significant portion of the data was taken at high instantaneous
luminosity, where more than one interaction per beam crossing was
probable.  When an event had more than one reconstructed vertex, the
quantity $S_T=|\Sigma \vec{p}_T^{\;\text{jet}}|$ was defined for the
two vertices that had the largest numbers of associated tracks, and
the vertex with the smallest $S_T$ was used for calculating all
kinematic variables~\cite{d0_jets_prd,thesis}.  To preserve the
pseudo-projective nature of the D\O\ calorimeter, the vertex
$z$-position was required to be within $50$ cm of the center of the
detector. This requirement rejected $(10.6\pm0.1)\%$ of the events,
independent of jet transverse momentum.

Isolated noisy calorimeter cells were suppressed with online and
offline algorithms~\cite{thesis}.  Background introduced by electrons,
photons, detector noise, and accelerator losses that mimicked jets
were eliminated with jet quality cuts. The efficiency of the jet
selection is approximately $99.5\%$ and nearly independent of jet
$p_T$.  The imbalance in transverse momentum, ``missing transverse
energy,'' was calculated from the vector sum of the $E_{x,y}$ values
in all cells of the calorimeter.  Background from cosmic rays or
incorrectly vertexed events was eliminated by requiring the missing
transverse energy in each event to be less than $70\%$ of the \pt of
the leading jet.  This criterion caused essentially no loss in
efficiency.

The D\O\ jet momentum calibration~\cite{rob_prd}, applied on a
jet-by-jet basis, corrects on average the reconstructed calorimeter
jet \pt to that of the final-state particles in the jet.  The
correction accounts for contribution from background from spectator
partons (the ``underlying-event,'' determined from minimum-bias
events), additional interactions, pileup from previous $p\overline{p}$
crossings, noise from uranium radioactivity, detector
non-uniformities, and for the global response of the detector to
hadronic jets. Unlike the cone algorithm, the \kt algorithm does not
require additional corrections for showering in the
calorimeter~\cite{rob_prd}.  For $|\eta|<0.5$, the mean total
multiplicative correction factor to an observed \pt of $100$ ($400$)
GeV is $1.094\pm0.015$ ($1.067\pm0.020$).

\begin{figure}[t]
\begin{center}
\mbox{}
\epsfxsize = 3.375in \epsfbox{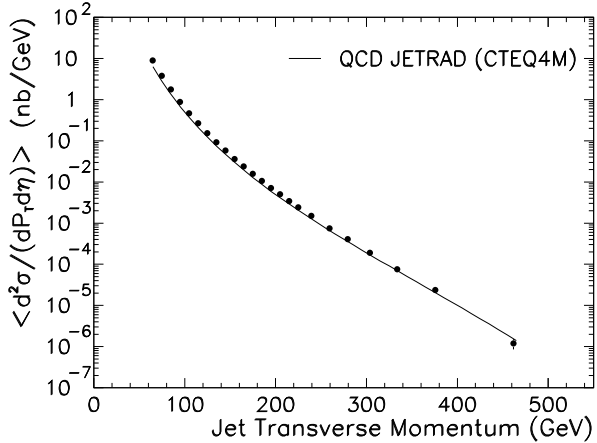}
\end{center}
\vskip 0.3cm
\caption{The central ($|\eta|<0.5$) inclusive jet cross 
section obtained with the \kt algorithm at $\sqrt{s}=1.8$ TeV. 
Only statistical errors are included. The solid line shows a 
prediction from NLO pQCD.}
\label{fig:xsection}
\end{figure}

The inclusive jet cross section for $|\eta|<0.5$ was calculated over
four ranges of transverse momentum, each using data from only a single
trigger threshold. The more restrictive trigger was used as soon as it
became fully efficient.  The average differential cross section for
each \pt bin, $d^2\sigma/(dp_T d\eta)$, was measured as $N/(\Delta\eta
\Delta p_T \epsilon L)$, where $\Delta\eta$ and $\Delta p_T$ are the
$\eta$ and $p_T$ bin sizes, $N$ is the number of jets observed in that
bin, $\epsilon$ is the overall efficiency for jet and event selection,
and $L$ represents the integrated luminosity of the data sample.

\begin{table}[t]
\squeezetable
\begin{center}
\begin{tabular}{cccc}
 \pt Bin & Plotted & Cross Sec. $\pm$ Stat. 
& Systematic \\
  (GeV) &  \pt (GeV) & (nb/GeV) & Uncer. ($\%$) \\
\hline\hline
 $ 60 -  70$ & $ 64.6$ & $ (8.94\pm 0.06)\times 10^{0\,\:\:\:} 	
	$ & $-13$, $+14$ \\
 $ 70 -  80$ & $ 74.6$ & $ (3.78\pm 0.04)\times 10^{0\,\:\:\:} 
	$ & $-13$, $+14$ \\
 $ 80 -  90$ & $ 84.7$ & $ (1.77\pm 0.02)\times 10^{0\,\:\:\:} 
 	$ & $-13$, $+14$ \\
 $ 90 - 100$ & $ 94.7$ & $ (8.86\pm 0.25)\times 10^{-1}$ & $-13$, $+14$ \\
 $100 - 110$ & $104.7$ & $ (4.68\pm 0.04)\times 10^{-1}$ & $-14$, $+14$ \\
 $110 - 120$ & $114.7$ & $ (2.68\pm 0.03)\times 10^{-1}$ & $-14$, $+14$ \\
 $120 - 130$ & $124.8$ & $ (1.53\pm 0.02)\times 10^{-1}$ & $-14$, $+14$ \\
 $130 - 140$ & $134.8$ & $ (9.19\pm 0.16)\times 10^{-2}$ & $-14$, $+14$ \\
 $140 - 150$ & $144.8$ & $ (5.77\pm 0.12)\times 10^{-2}$ & $-14$, $+14$ \\
 $150 - 160$ & $154.8$ & $ (3.57\pm 0.03)\times 10^{-2}$ & $-15$, $+14$ \\
 $160 - 170$ & $164.8$ & $ (2.39\pm 0.02)\times 10^{-2}$ & $-15$, $+14$ \\
 $170 - 180$ & $174.8$ & $ (1.56\pm 0.02)\times 10^{-2}$ & $-15$, $+14$ \\
 $180 - 190$ & $184.8$ & $ (1.05\pm 0.02)\times 10^{-2}$ & $-15$, $+14$ \\
 $190 - 200$ & $194.8$ & $ (7.14\pm 0.13)\times 10^{-3}$ & $-16$, $+15$ \\
 $200 - 210$ & $204.8$ & $ (4.99\pm 0.08)\times 10^{-3}$ & $-16$, $+15$ \\
 $210 - 220$ & $214.8$ & $ (3.45\pm 0.07)\times 10^{-3}$ & $-16$, $+15$ \\
 $220 - 230$ & $224.8$ & $ (2.43\pm 0.06)\times 10^{-3}$ & $-16$, $+15$ \\
 $230 - 250$ & $239.4$ & $ (1.50\pm 0.03)\times 10^{-3}$ & $-17$, $+16$ \\
 $250 - 270$ & $259.4$ & $ (7.52\pm 0.23)\times 10^{-4}$ & $-17$, $+16$ \\
 $270 - 290$ & $279.5$ & $ (4.07\pm 0.17)\times 10^{-4}$ & $-18$, $+17$ \\
 $290 - 320$ & $303.8$ & $ (1.93\pm 0.09)\times 10^{-4}$ & $-18$, $+18$ \\
 $320 - 350$ & $333.9$ & $ (7.61\pm 0.59)\times 10^{-5}$ & $-19$, $+19$ \\
 $350 - 410$ & $375.8$ & $ (2.36\pm 0.23)\times 10^{-5}$ & $-20$, $+21$ \\ 
 $410 - 560$ & $461.8$ & $ (1.18\pm 0.33)\times 10^{-6}$ & $-23$, $+27$ \\
\end{tabular}
\end{center}
\vspace*{-.2cm}
\caption{Inclusive jet cross section of jets reconstructed using the
\kt algorithm in the central pseudorapidity region ($|\eta|<0.5$). 
}
\label{table:xsec}
\end{table}

\begin{figure}[t]
\begin{center}
\mbox{}
\epsfxsize =3.375in \epsfbox{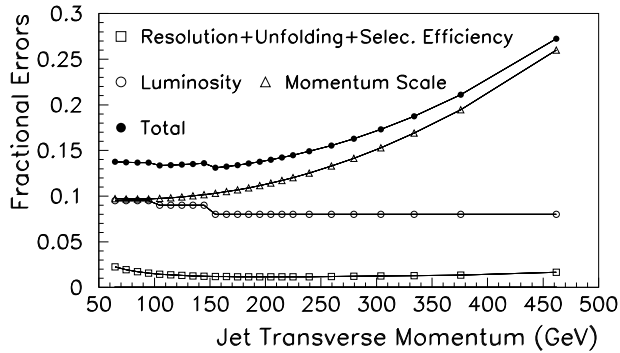}
\end{center}
\vskip -0.cm
\caption{Fractional experimental uncertainties on the cross section. 
The discontinuities in the luminosity uncertainty
are related to the use of different triggers~[11].}
\label{fig:fractional_errors}
\end{figure}

The measured cross section is distorted in \pt by the momentum
resolution of the D\O\ calorimeter.  The fractional momentum
resolution was determined from the imbalance in \pt in two-jet
events~\cite{thesis}.  Although the resolution in jet \pt is
essentially Gaussian, the steepness of the \pt\ spectrum shifts the
observed cross section to larger values.  At $100$ ($400$) GeV, the
fractional resolution is $0.061\pm0.006$ ($0.039\pm0.003$).  The
distortion in the cross section due to the resolution was corrected by
assuming an ansatz function, $A\,p_T^{-B} (1-2\,p_T/\sqrt{s}\,)^C$,
smearing it with the measured resolution, and fitting the parameters
$A$, $B$ and $C$ so as to best describe the observed cross section.
The bin-to-bin ratio of the original ansatz to the smeared one was
used to remove the distortion due to resolution. The unsmearing
correction reduces the observed cross section by $(5.7\pm1)\%$
($(6.1\pm1)\%$) at $100$ ($400$) GeV.

\begin{figure}[t]
\begin{center}
\mbox{}
\epsfxsize = 3.375in \epsfbox{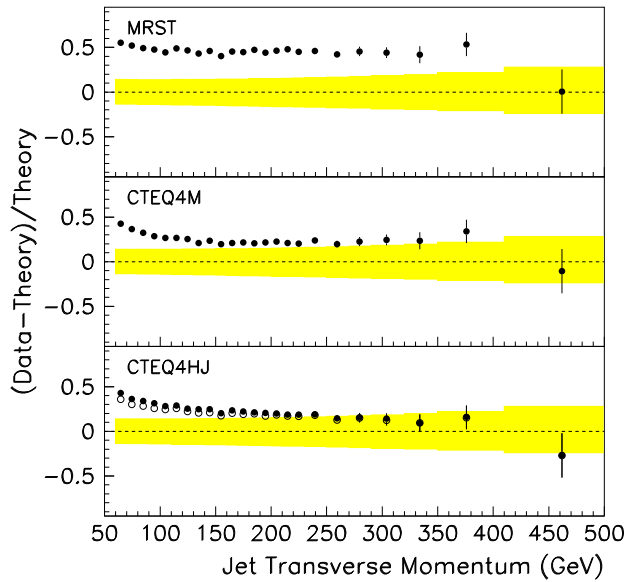}
\end{center}
\vskip -0.55cm
\caption{Difference between data and {\sc jetrad} pQCD, normalized to the
predictions. The shaded bands represent the total systematic uncertainty.
In the bottom plot a {\sc herwig} hadronization contribution has been 
added to the prediction (open circles).}
\label{fig:dtt}
\end{figure}

\begin{figure}[t]
\begin{center}
\mbox{}
\epsfxsize = 3.375in \epsfbox{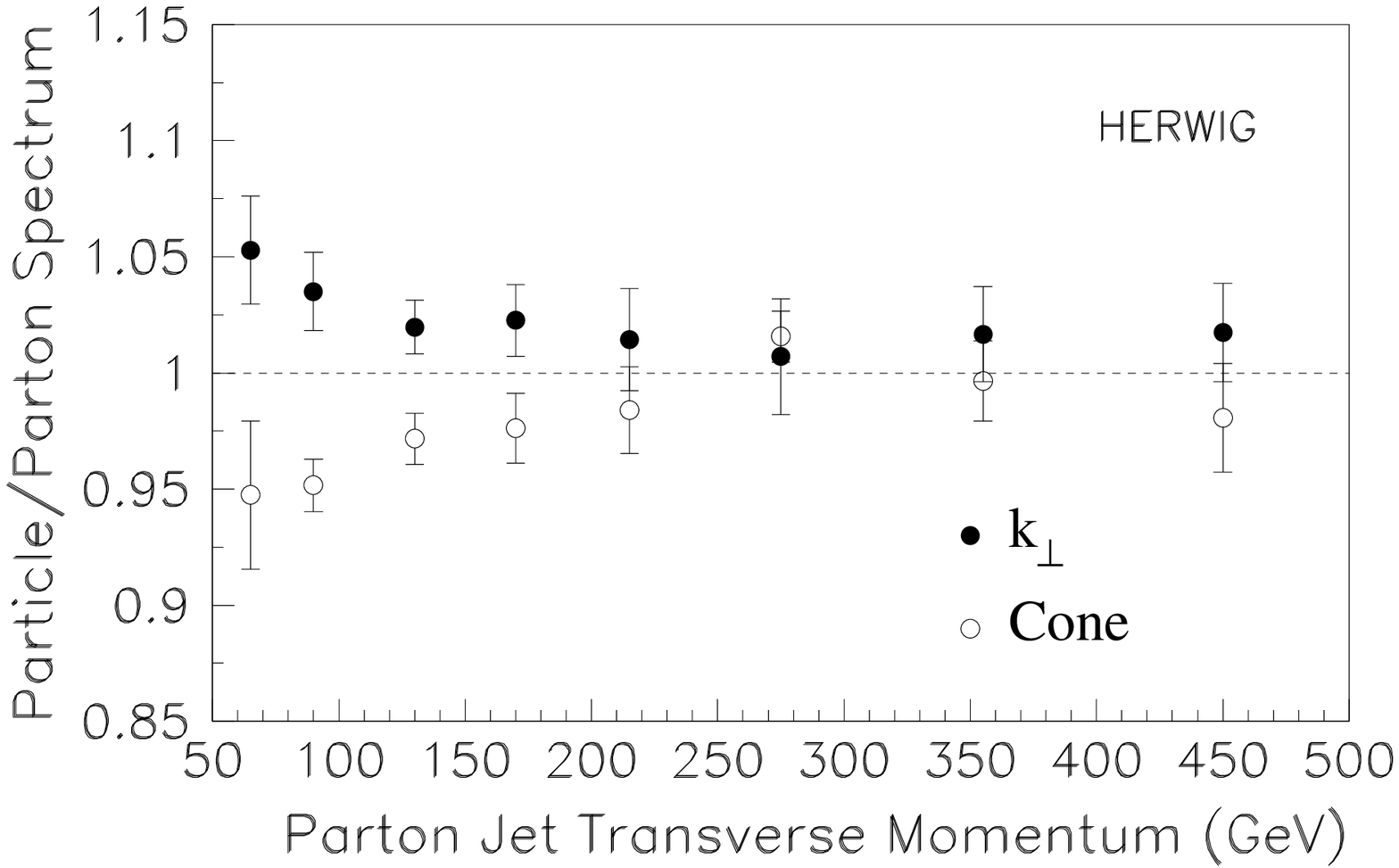}
\end{center}
\vskip .0cm
\caption{Ratio of particle-level over parton-level {\sc herwig} \pt spectra 
for jets, as a function of the parton jet transverse momentum.
}
\label{fig:kt_cone_herwig}
\end{figure}

The final, fully corrected cross section for $|\eta|<0.5$ is shown in
Fig.~\ref{fig:xsection}, along with the statistical uncertainties.
Listed in Table~\ref{table:xsec} are the \pt range, the best \pt bin
centroid, the cross section, and uncertainties in each bin.  The
systematic uncertainties include contributions from jet and event
selection, unsmearing, luminosity, and the uncertainty in the momentum
scale, which dominates at all transverse momenta.  The fractional
uncertainties for the different components are plotted in
Fig.~\ref{fig:fractional_errors} as a function of the jet transverse
momentum.

The results are compared to the pQCD NLO prediction from {\sc
jetrad}~\cite{jetrad}, with the renormalization and factorization
scales set to $p_T^{\text{max}}/2$, where $p_T^{\text{max}}$ refers to
the \pt of the leading jet in an event. The comparisons are made
using parametrizations of the parton distribution functions (PDFs) of
the CTEQ~\cite{cteq} and MRST~\cite{mrs} families.
Figure~\ref{fig:dtt} shows the ratios of (data-theory)/theory.  The
predictions lie below the data by about $50\%$ at the lowest \pt and
by $(10-20)\%$ for $p_T>200$ GeV.  To quantify the comparison in
Fig.~\ref{fig:dtt}, the fractional systematic uncertainties are
multiplied by the predicted cross section, and a $\chi^2$ comparison,
using the full correlation matrix, is carried out~\cite{d0_jets_prd}.
The results are shown in Table~\ref{table:chi2}.  Though the agreement
is reasonable ($\chi^2/\rm{dof}$ ranges from $1.56$ to $1.12$, the
probabilities from $4$ to $31\%$), the differences in normalization
and shape, especially at low {\it p}$_{T}$, are quite large.  The
points at low \pt\ have the highest impact on the $\chi^2$.  If the
first four data points are not used in the $\chi^2$ comparison, the
probability increases from $29\%$ to $77\%$ when using the CTEQ4HJ
PDF.

\begin{table}[t]
\squeezetable
\begin{center}
{\footnotesize
\begin{tabular}{l c c c}
    PDF &  \raisebox{-.0cm}{$\chi^2$} 
        &  \raisebox{-.0cm}{$\chi^2/dof$}  
        &  Probability ($\%$) \\
\hline
MRST              & 26.8 & 1.12 & 31 \\
MRSTg$\uparrow$   & 33.1 & 1.38 & 10 \\
MRSTg$\downarrow$ & 28.2 & 1.17 & 25 \\
CTEQ3M            & 37.5 & 1.56 & 4  \\
CTEQ4M            & 31.2 & 1.30 & 15 \\
CTEQ4HJ           & 27.2 & 1.13 & 29 \\
MRST+Hadroniz.    & 24.0 & 1.00 & 46 \\
CTEQ4HJ+Hadroniz. & 24.3 & 1.01 & 44 \\
\end{tabular}
}
\end{center}
\vspace*{-.4cm}
\caption{
$\chi^2$ comparison ($24$ degrees of freedom) between {\sc jetrad}, with
renormalization and factorization scales set to $p_T^{\text{max}}/2$, and data
for various PDFs.  The last entries include a hadronization correction
obtained from {\sc herwig} (see text).
}
\label{table:chi2}
\end{table}

While the NLO predictions for the inclusive cross section for
\kt ($D=1.0$) and cone jets ($R=0.7$, \Rsep$=1.3$
in the same $|\eta|<0.5$ interval are within $1\%$ of each other for
the \pt\ range of this analysis~\cite{thesis}, the measured cross
section using \kt is $37\%$ ($16\%$) higher than the previously
reported cross section using the cone algorithm~\cite{d0levan} at $60$
($200$) GeV. This difference in the cross sections is consistent with
the measured difference in \pt\ for cone jets matched in $\eta-\phi$
space to \kt\ jets. \kt\ jets were found to encompass $7\%$ ($3\%$)
more transverse energy at $60$ ($200$) GeV than cone
jets~\cite{rob_prd,thesis}.

The effect of final-state hadronization on reconstructed energy, which
might account for the discrepancy between the observed cross section
using \kti\ and the NLO predictions at low $p_T$, and also for the
difference between the \2 and cone results, was studied using {\sc
herwig} (version $5.9$)~\cite{herwig} simulations.
Figure~\ref{fig:kt_cone_herwig} shows the ratio of
\pt spectra for particle-level to parton-level jets, for both the \kti\ 
and cone algorithms.  Particle cone jets, reconstructed from final
state particles (after hadronization), have less \pt than the parton
jets (before hadronization), because of energy loss outside the cone.
In contrast, \kt particle jets are more energetic than their
progenitors at the parton level, due to the merging of nearby partons
into a single particle jet.  Including the hadronization effect
derived from {\sc herwig} in the NLO {\sc jetrad} prediction improves
the $\chi^2$ probability from $29\%$ to $44\%$ ($31\%$ to $46\%$) when
using the CTEQ4HJ (MRST) PDF.  We have also investigated the
sensitivity of the measurement to the modeling of the background from
spectator partons through the use of minimum bias events, and found
that it has a small effect on the cross section: at low $p_T$, where
the sensitivity is the largest, an increase of as much as $50\%$ in
the underlying event correction decreases the cross section by less
than $6\%$.

In conclusion, we have presented the first measurement in
proton-antiproton collisions at $\sqrt{s}=1.8$ TeV of the inclusive
jet cross section using the \kt algorithm.  Quantitative tests show
reasonable agreement between data and NLO pQCD predictions, except at
low $p_T$ where the agreement is marginal.  The degree of agreement
can be slightly improved by incorporating a hadronization contribution
of the kind predicted by {\sc herwig}.


%
We thank the staffs at Fermilab and collaborating institutions, 
and acknowledge support from the 
Department of Energy and National Science Foundation (USA),  
Commissariat  \` a L'Energie Atomique and 
CNRS/Institut National de Physique Nucl\'eaire et 
de Physique des Particules (France), 
Ministry for Science and Technology and Ministry for Atomic 
   Energy (Russia),
CAPES and CNPq (Brazil),
Departments of Atomic Energy and Science and Education (India),
Colciencias (Colombia),
CONACyT (Mexico),
Ministry of Education and KOSEF (Korea),
CONICET and UBACyT (Argentina),
The Foundation for Fundamental Research on Matter (The Netherlands),
PPARC (United Kingdom),
Ministry of Education (Czech Republic),
and the A.P.~Sloan Foundation.
%




\end{document}